# A Generalized Switched-Capacitor Modular Multilevel Inverter Topology for Multiphase Electrical Machines with Capacitor-Voltage Self-Balancing Capability

**Farzad Iraji**
Department of Electrical and Computer Engineering
Technische Universität Kaiserslautern
Kaiserslautern, DE 67663
Iraji@eit.uni-kl.de

**Nima Tashakor**
Department of Electrical and Computer Engineering
Technische Universität Kaiserslautern
Kaiserslautern, DE 67663
Tashakor@eit.uni-kl.de

**Stefan Goetz**
Electrical and Computer Engineering
Duke University
Durham, NC 27710
Stefan.goetz@duke.edu

April 11, 2022

## Abstract

Recent research on multilevel inverters shows exciting properties, including the potential to generate multiple output voltages and integrated voltage boosting. However, most presented inverter topologies have a restricted number of output voltage levels and limited output voltage boosting ratio. In addition, balancing the voltage of capacitors in multilevel converters is very important and should be considered in the topology or control method. This paper describes a generalized switched-capacitor circuit topology for multilevel dc-to-ac inverters that can be developed for the desired output voltage levels. Also, the maximum ac output voltage can vary from values much lower than the dc input voltage to several times of it depending on the design requirement. High-voltage dc input and ac output could be handled using low-voltage capacitors, which substantially decreases overall cost and volume. The proposed topology further allows for easy expansion through stackable circuits for multiple load phases. It has an inherent capacity for balancing the voltage of capacitors. To validate the feasibility and practicality of this concept, we provide circuit descriptions, control strategies, design recommendations, and pertinent simulation findings for the suggested inverter topology.

## 1 Introduction

The interest in electrical machines is multiplied with the added emphasis on greener transportation such as electric and hybrid vehicles, electric ship propulsion, and electric locomotive traction. Conventional electric systems based on three-phase machinery are not necessarily the best solutions for these applications. Indeed, multiphase electrical machines (> 3) can offer various potential advantages over their three-phase equivalent, such as better power distribution over more phases, lower torque ripple, and better fault tolerance [1]. Regardless of the number of phases, electrical machines require a dc–ac inverter for proper parameter controlling.

There are various types of dc–ac inverters such as conventional H-bridge, cascaded H-bridge [2–4], Z-source [5–7], multilevel [8–10], and modular multilevel [11–17]. In recent years, multilevel inverters (MLI) have been researched more due to their advantages over conventional three-level PWM converters, such as smaller output harmonics, lower switches power ratings, high voltage capability, and lower $dv/dt$ stress on switches [18]. The conventional MLI topologies could be categorized into three major groups known as *classical topologies* formed by cascaded H-bridge (CHB) [19], neutral-point clamped (NPC) [20], and flying-capacitor (FC) MLI [21].

The classical topologies' complexity and cost will increase significantly in higher voltage levels, mainly due to the exponential growth of the components such as switches, gate drivers, and capacitors. Therefore, new topologies with a reduced number of switches are necessary. Saeidabadi et al. propose a three-phase hybrid MLI consisting of three single-phase H-bridge inverters, a three-phase H-bridge inverter, and auxiliary modules for increasing the number of output voltage levels with a low number of components [22]. Karasani et al. likewise present an altered three-phase h-bridge inverter with fewer components where they practically incorporate a t-type structure into the modules [16]. Batschauer et al., on the other hand, describe a three-phase hybrid MLI for medium-voltage applications which employs a three-phase voltage-source inverter (VSI) and series-connected half-bridge modules [23]. Ali et al. likewise adopt a module structure that uses t-type features to form a single-phase asymmetrical MLI with fewer components for distribution systems [24]. Babaei et al. formalized and generalized the principle for MLI with H-bridge circuits [25]. However, the number of individual dc sources is not decreased. On the contrary, some topologies such as the proposed topology in [26] use even more components compared to classical topologies.

It should be mentioned that the reliability of MLI is jeopardized by capacitor voltage imbalance. One of the critical hurdles for MLI expansion is keeping the sub-module (SM) capacitor voltages balanced at their nominal values [27]. MMCs need a technique to actively balance the voltage of capacitors within a tolerable range through additional parallel connections or software-based balancing loops [28–30] or use modified topologies that mix MLI and switched-capacitor or switch-inductor functionality for sensorless operation [31–33]. The most common approach for balancing modules' voltages is sorting [34, 35]; when the arm current is negative, the highest-voltage modules are placed in series first, while the lowest-voltage modules are inserted in series last, and vice versa [36–40]. The fundamental issue of cell-sorting approaches is their need to either measure or estimate module voltages, which increases cost and complicates algorithms while also lowering reliability [41]. New chopper-circuit topologies for dc capacitor voltage balancing in diode-clamped MLI (DCMLIs) and advanced control features are known in the literature [42–44]. Hasegawa et al. present a voltage balancing circuit with two unidirectional choppers and a single coupled inductor with two electrically isolated windings with phase-shift control [45]. As an alternative, a parallel-connected diode-clamped MMC topology with two parallel-connected clusters in each arm may promote the concept of self-balancing [46]. The capacitor voltages can be balanced automatically in this design without the need for any balancing control techniques. However, it increases the number of components. In [43], a single string diode-clamped MMC topology is proposed, which can be considered as the simplest diode-clamped self-balancing topology. However, the self-balancing might be at the cost of slightly lower efficiency in highly imbalanced systems.

Another important stream of research of recent decades in power electronics, though typically for dc/dc conversion and thus a different application, are switched capacitor circuits [47]. Switched-capacitor converters use series–parallel reconfiguration of capacitors to adjust voltages and are therefore very similar to cascaded converters and MLIs. Prior research presented topologies that can exploit switched-capacitor features and modular multilevel converters at the same time, sometimes even through the same transistors, to generate dc or ac [48, 49]. However, these circuits operate switched-capacitor modes on top of modular multilevel functionality between modules, not within. In principle, the structure of cascaded bridges, H bridges as well as chopper circuits, with their internal dc design, however, appears ideal for the combination with switched-capacitor circuits to increase the functionality and flexibility. In principle, nesting would be a reasonable solution and is known in general [50]. Existing approaches, however, require a large number of transistors and/or capacitors and sometimes even involve diodes with their performance issues [51, 52].

This paper proposes a switched-capacitor multilevel inverter with a generalized circuit topology that may output various voltage levels. SM can be used alone; however, a novel configuration for linking SMs is proposed. The MMI is appropriate for multiphase systems due to the suggested topology of SM and configuration design. The proposed architecture is modular, making extension simple. It has the ability to balance capacitor voltage and can increase or decrease output voltage in comparison to the input dc voltage. Furthermore, it can generate high-voltage AC using low-voltage capacitors, thus lowering the overall cost and volume.

## 2 Proposed MMI Topology

Fig. 1 shows the most frequently used basic SMs of conventional multilevel converters. Each SM has at least one energy storage device such as capacitors, as well as a number of semiconductor devices, such IGBTs, FETs, and diodes. For voltage sharing, SMs are linked in series, resulting in two output terminals, each coupled to a nearby SM. Asymmetrical half-bridge topology is shown in Fig. 1(a) in which the two switches $S_{i1}$ and $S_{i2}$ are complementary that means if $S_{i1}$ is $ON/OFF$ then $S_{i2}$ should be $OFF/ON$. If $S_{i1}$ is $ON$ and $S_{i2}$ is $OFF$, then the SM will supply a positive voltage, otherwise will be bypassed with zero output voltage. The same concept applies to H-bridge SM shown in Fig. 1(b). Typically, upper switches ($S_{i1}$ and $S_{i3}$) and lower switches ($S_{i2}$ and $S_{i4}$) are complementary. H-bridge SM should produce a positive or negative voltage when diagonal switches $S_{i1}$ and $S_{i4}$ (or $S_{i2}$ and $S_{i3}$) turn $ON/OFF$ simultaneously; otherwise zero.

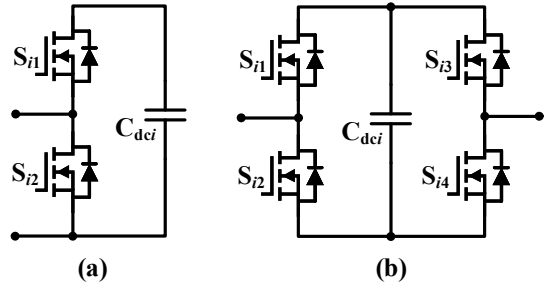

Figure 1: Conventional multilevel converters' basic SM, a) Asymmetrical half-bridge, b) H-bridge.

The proposed multiphase MMI includes two basic parts. The first part is the main circuit, which comprises several modular SM connected in series. The control circuit, which is divided into two sections for load management and capacitor balancing, is the second part and is described in Section 3. Furthermore, for simplicity's sake, the proposed topology is presented in five-level output first, followed by the generalized topology. It is assumed in all SM descriptions in the following that $V_{C_1} = V_{C_2} = ... = V_{C_n} = \frac{V_{DC}}{n}$ (see Fig. 2). This equation is covered in more detail in Section 3.

## 2.1 Five-Level SM Description

The proposed SM is illustrated in Fig. 2(a). Seven power switches ($S_{i1}$ to $S_{i7}$) and two dc-link capacitors ($C_{i1}$ and $C_{i2}$) are used in the suggested design. Upper ($S_{i1}$ and $S_{i3}$) and lower switches ($S_{i2}$ and $S_{i4}$) are complementary. Between $S_{i1}$ to $S_{i4}$, either the upper or lower switches should have internal antiparallel diodes, and the other ones may or may not have the diode, but the $S_{i7}$ must not. Five different current paths are presented to provide all the individual output voltage levels, as indicated in Fig. 2(b)–(g). Following is a description of how the proposed SM works in each of these five states. Table 1 summarizes the various voltage levels.

### 2.1.1 Zero-level paths

In both half cycles of the sinusoidal phase voltage, two $ON$ state power switches contribute to the zero level of the phase output voltage ($V_i = 0$). As shown in Fig. 2(b)-(c), zero-level could be implemented using two different paths, either ($S_{i1}$ and $S_{i3}$) or ($S_{i2}$ and $S_{i4}$). As can be seen, $C_{i1}$ and $C_{i2}$ are disconnected from the load, while they could be paralleled to balance and charged up to the input DC source voltage ($V_{C_{i1}} = V_{C_{i2}} = V_{DC}$), ignoring the tiny voltage drop on the switches.

### 2.1.2 First positive level ($+V_{C_{i1}}$ mode)

The first positive level of the inverter output voltage is produced by $S_{i2}$, $S_{i3}$, $S_{i6}$, and $S_{i7}$ as indicated in Fig. 2(d) with a green line ($V_i = V_{C_{i1}}$). Besides, $C_{i2}$ is charged up to $V_{C_{i1}}$ by the connection provided by the $ON$ state switches, so $V_{C_{i1}} = V_{C_{i2}}$.

### 2.1.3 Second positive level ($+2V_{C_{i1}}$ mode)

Using $C_{i1}$ and the charged $C_{i2}$ from the first positive level, the second positive level of the output voltage ($V_i = V_{C_{i1}} + V_{C_{i2}}$) is produced. Since the second positive level always comes after the first positive-level, $V_{C_{i1}} = +V_{C_{i2}}$, then $V_i = 2V_{C_{i1}}$. For this purpose, three power switches $S_{i2}$, $S_{i3}$, and $S_{i5}$ should be turned $ON$, the current path is indicated as a green line in Fig. 2(e).

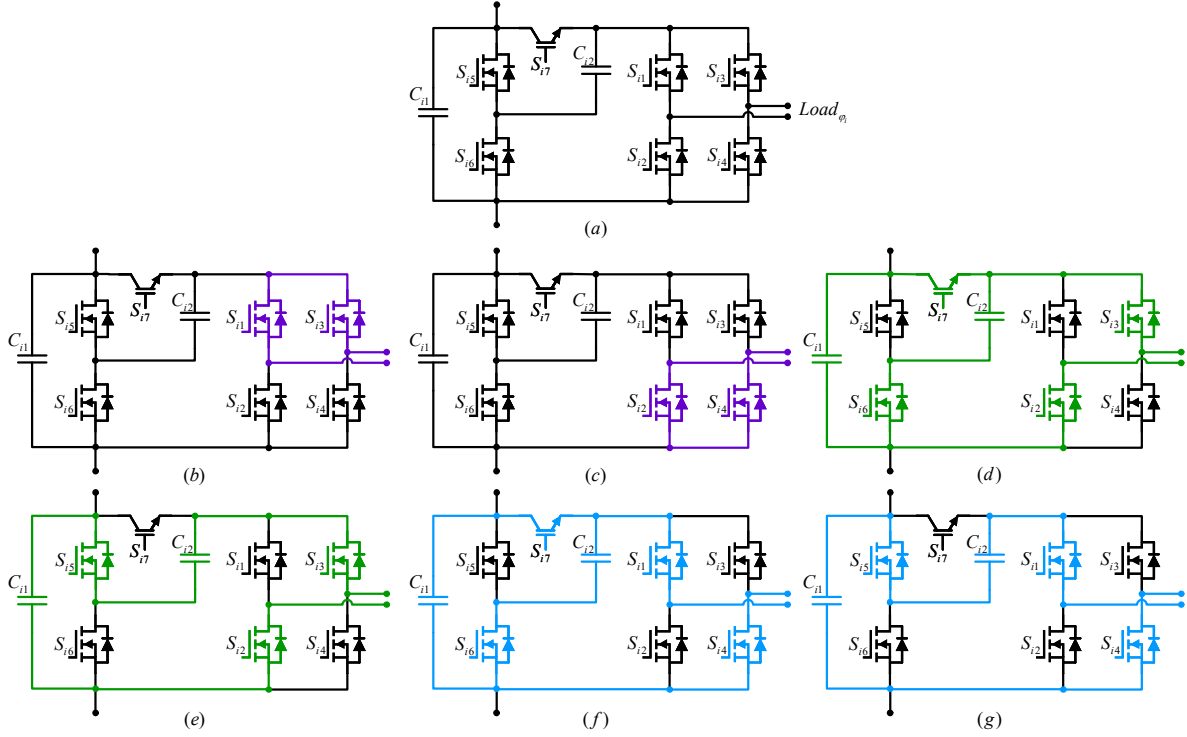

Figure 2: The proposed SM topology and current paths, a) the SM illustration, b) the first zero-level path, c) the second zero-level path, d) the first positive-level($+V_{C_{i1}}$), e) the second positive-level ($+2V_{C_{i1}}$), f) the first negative-level ($-V_{C_{i1}}$), g) the second negative-level ($-2V_{C_{i1}}$).

Table 1: Proposed five-level SM operation states

| Switches Conditions | | | | | | | Output |
|---|---|---|---|---|---|---|---|
| $S_{i1}$ | $S_{i2}$ | $S_{i3}$ | $S_{i4}$ | $S_{i5}$ | $S_{i6}$ | $S_{i7}$ | $V_i$ |
| $ON$ | $OFF$ | $ON$ | $OFF$ | $OFF$ | $OFF$ | $OFF$ | 0 |
| $OFF$ | $ON$ | $OFF$ | $ON$ | $OFF$ | $OFF$ | $OFF$ | 0 |
| $OFF$ | $ON$ | $ON$ | $OFF$ | $OFF$ | $ON$ | $ON$ | $+V_{C_{i1}}$ |
| $OFF$ | $ON$ | $ON$ | $OFF$ | $ON$ | $OFF$ | $OFF$ | $+2V_{C_{i1}}$ |
| $ON$ | $OFF$ | $OFF$ | $ON$ | $OFF$ | $ON$ | $ON$ | $-V_{C_{i1}}$ |
| $ON$ | $OFF$ | $OFF$ | $ON$ | $ON$ | $OFF$ | $OFF$ | $-2V_{C_{i1}}$ |

### 2.1.4 First negative level ($-V_{C_{i1}}$ mode)

Switches $S_{i1}$, $S_{i4}$, $S_{i6}$, and $S_{i7}$ should be switched $ON$ to provide the first negative-level output voltage ($V_i = -V_{C_{i1}}$). This is shown in Fig. 2(f) as a blue line. The $ON$ state switches $S_{i6}$ and $S_{i7}$ create a path for $C_{i2}$ to be charged up to $V_{C_{i1}}$ again, the final state is $V_{C_{i1}} = V_{C_{i2}}$.

### 2.1.5 Second negative level ($-2V_{C_{i1}}$ mode)

Finally, as illustrated in Fig. 2(g), the second negative level of the output voltage is generated while three power switches $S_{i1}$, $S_{i4}$, and $S_{i5}$ are switched $ON$. In this condition the output voltage is $V_i = -(V_{C_{i1}} + V_{C_{i2}})$. Since the second negative level always should be produced after the first negative level $V_{C_{i1}} = V_{C_{i2}}$, then $V_i = -2V_{C_{i1}}$.

## 2.2 Generalized SM Topology

The proposed SM could be expanded to produce $N_L \geq 3$ where $N_L$ is the number of output levels. As illustrated in Fig. 3, a generalized topology may be developed in which each two more voltage level requires three switches and one capacitor to produce. The extra switches may or may not have internal antiparallel diodes, but the ones shown in Fig. 3 without one must not. The generalized MMI's operating concept is the same as that of the proposed five-level MMI. All the capacitors are self-balanced at the input dc-voltage value, eliminating the need for voltage sensors or a complex control platform. The number of new components to express the proposed topology's generalization capabilities may be written as

$$N_S = \frac{3N_L - 1}{2}, \tag{1}$$

$$N_C = \frac{N_L - 1}{2}, \tag{2}$$

where $N_C$ is the number of required capacitors and $N_S$ is the number of total required switches. As previously stated, alternative switches might be used in the SM architecture in terms of antiparallel diode in switches. The numbers of each kind are

$$N_{NA} = \frac{N_L - 3}{2}, \tag{3}$$

$$N_{WA} = 2, \tag{4}$$

$$N_{OW} = N_L - 1, \tag{5}$$

where $N_{NA}$ is the number of switches that must not have an antiparallel diode, $N_{WA}$ is the number of switches that must have the diode and $N_{OW}$ is the number of switches that might be of any type.

## 2.3 System Configuration

The traditional method for connecting SMs in MLIs is to connect them in parallel with the dc source. The input voltages of SMs are similar in this way. However, in this study, a connecting method is described to improve voltage compatibility, significantly, which means that low-voltage capacitors could be employed with a high-voltage input dc source and the output voltage could be anywhere from a fraction of $V_{DC}$ to several times higher. Connecting $n$ SMs in series using the connections which are on top and also the bottom of each SM (see Fig. 3) would result in a multiport multiphase MMI, where $n$ is the number of output phases. The proposed topology is shown in Fig. 4, which for an $n$-phase system, $n \times N_C$ capacitors requires $n \times N_S$ switches and just a single DC voltage source. Since each load phase is supplied directly from an SM, the phase voltage could be one of $N_L$ voltage levels that each SM could produce.

As the SMs are identical, $C_{11} = C_{21} = ... = C_{n1}$, considering the connection of capacitors that is series, the $V_{C_{i1}}$ could be concluded as follows:

$$C_{11} = C_{21} = ... = C_{n1} \rightarrow V_{C_{11}} = V_{C_{21}} = ... = V_{C_{n1}} \tag{6}$$

$$\begin{cases} V_{C_{11}} = V_{C_{21}} = ... = V_{C_{n1}} \\ V_{DC} = \sum_{i=1}^{n} V_{C_{i1}} \end{cases} \tag{7}$$
$$\rightarrow V_{C_{11}} = V_{C_{21}} = ... = V_{C_{n1}} = \frac{V_{DC}}{n}$$

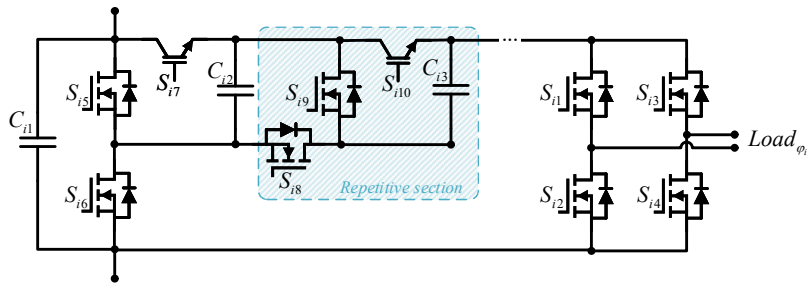

Figure 3: Proposed generalized SM topology.

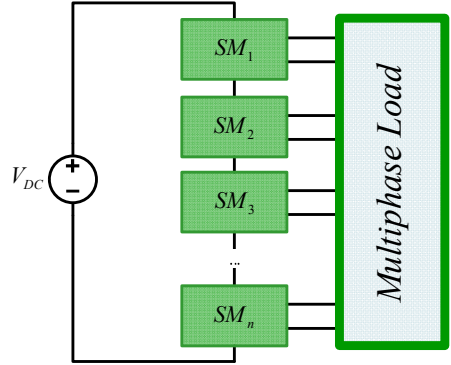

Figure 4: Proposed topology of multiport MMI for multiphase loads.

In comparison to $V_{C_{i1}}$, phase voltages could be boosted due to the SMs' generalized multilevel capability. The different voltage levels that each phase could reach follow

$$V_i = \left\{ \pm \frac{kV_{DC}}{n} | k \in \mathbb{Z}, 0 < 2k+1 \leq N_L \right\}. \tag{8}$$

Eq. (7) shows that the proposed topology has an intrinsic voltage reduction capability on capacitors, which reduces the cost and capacitors' volume significantly. Besides, according to Eq. (8), the phase voltage could be boosted to voltages beyond the dc source voltage if

$$N_L > 2n + 1. \tag{9}$$

Independent of $N_L$, the capacitors' nominal voltage and drain–source voltage of switches $S_{i5}$ to $S_{in}$ should be $V_{DC}/n$. Switches $S_{i1}$ to $S_{i4}$, however, should withstand the maximum voltage level produced which is $((N_L - 1)V_{DC}/2n)$ over their drain–source.

## 3 Proposed Control System

Since each phase voltage is directly related to the voltage of $C_{i1}$, the specific value of the voltage is important. If the capacitors are identical and other components are also ideal, as proven in Eq. (7), the voltages of all the capacitors would be the same. However, several aspects could influence the behavior of such a system, such as capacitors manufacturing differences, aging, and even various load transients on each phase. Since the proposed MMI's output capabilities all depend on balanced capacitor voltages, this is an essential type of situation to consider. No extra control system or component is required to balance the capacitors in the presented inverter for multiphase machines. Electromagnetic coupling exists between the windings of electrical devices according to the motor wiring. The mentioned coupling is considered in the development of the proposed topology to balance the capacitors automatically.

To describe the principle of automatic balancing, let us consider:

1. an MMI built with five-level SMs;
2. two SMs ($SM_H$ and $SM_L$) are in the first positive level state, then capacitors in each SM are parallel;
3. one of the capacitors' voltages ($C_H = C_{H1} + C_{H2}$) is by $\Delta V$ higher than average for any reason.

As a result of (3), another capacitor's voltage ($C_L = C_{L1} + C_{L2}$) is $\Delta V$ lower than the average voltage, ensuring that the total voltage remains constant at $V_{DC}$. Although the voltage decrement could be divided among multiple capacitors, the voltage division is ignored for the sake of simplicity. Then, the produced PWM voltage of $C_L$ is lower than average, which causes the regarded winding to act as a winding in a multiphase transformer with lower voltage which will drain power from the other windings. The drained power will charge the $C_L$ while simultaneously discharging the $C_H$. Charging the $C_L$ will increase the voltage of $C_{L2}$, but because there is no current path from $C_{L2}$ to $C_{L1}$, the extra charge will not be conducted to $C_{L1}$; however, no power will be drained from $C_{L1}$ either. On the other hand, the power is drained from $C_H$, which means $C_{H2}$ will be discharged—but this time there is a current path between $C_{H1}$ to $C_{H2}$ which will discharge $C_{H2}$, too. $C_{L1}$ not being discharged and $C_{H1}$ being discharged will make the voltage of $C_{L1}$ to increase. This process continues until $C_{H1} = C_{L1}$. This concept will act the same in systems with $N_L$-level SMs or distributed voltage decrement.

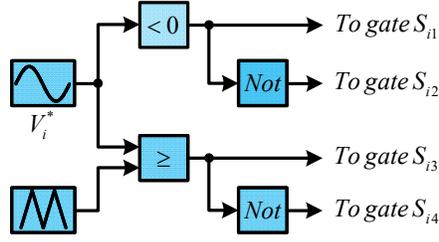

Figure 5: The proposed PWM method to produce proper sinusoidal output.

As a result, the presence of the windings' coupling will completely balance the system. The absence of coupling in a load, on the other hand, will not cause the voltages of capacitors to diverge. Assume the load is made up of unique resistors. If $C_H$ is $\Delta V$ higher for any reason, the load will drain $\Delta V/R = \Delta I$ more. This allows the system to work in a new operating point slightly different from the previous one. Unless the load is constant power, this notion is applicable to all balanced insensitive loads.

### 3.1 PWM for The Proposed MMI

We use a PWM approach to create a sinusoidal waveform in each output phase and the shifted reference voltage method to produce multiphase voltages' references. According to the switching concept described in Section 2, we propose the corresponding PWM method for each phase as shown in Fig. 5, where $V_i^*$ is the reference voltage for $i$th phase. The $V_i^*$ could be concluded using different load control strategies or even could be a simple sinusoidal waveform as

$$V_i^* = \sin\left(2\pi f t + \frac{2\pi i}{n}\right), \tag{10}$$

where $f$ is the frequency. As previously stated, upper and lower switches are complementary; this is evidenced by the control system (Fig. 5). The gate signal for switch $S_{i1}$ is generated by simply comparing the reference voltage with zero. However, the $S_{i3}$ is generated by comparing the $V_i^*$ with a high-frequency saw-tooth carrier. The switching frequencies of $S_{i1}$ and $S_{i2}$ are $2f$ which is relatively low but the switching frequency of $S_{i3}$ and $S_{i4}$ are equal to carrier frequency ($f_c$).

### 3.2 Proposed MMI Analysis

The harmonics and maximum output voltage calculations for the worst case ($N_L = 3$) are detailed below. The effect of carrier is disregarded in the calculations. For a general understanding of the aspects, this assumption is appropriate. Fourier series expansion for the voltage of phase $i$ could be expressed as

$$V_i = \frac{2V_{DC}}{n\pi} \sum_{m=2k+1}^{\infty} \left(\frac{\sin(m\omega t) - \sin(m(\omega t - 2\pi/3))}{m}\right). \tag{11}$$

This equation can be simplified as

$$\begin{aligned}
V_i &= \frac{4V_{DC}}{n\pi} \sum_{m=2k+1}^{\infty} \frac{1}{m} \cos\frac{m\pi}{6} \sin m\left(\omega t + \frac{\pi}{6}\right) \\
&= \frac{2\sqrt{3}V_{DC}}{n\pi} \left[\sin\left(\omega t + \frac{\pi}{6}\right) - \frac{1}{5}\sin 5\left(\omega t + \frac{\pi}{6}\right) - \frac{1}{7}\sin 7\left(\omega t + \frac{\pi}{6}\right) + \frac{1}{11}\sin 11\left(\omega t + \frac{\pi}{6}\right) + \frac{1}{13}\sin 13\left(\omega t + \frac{\pi}{6}\right)\right] \\
&\quad + \ldots .
\end{aligned} \tag{12}$$

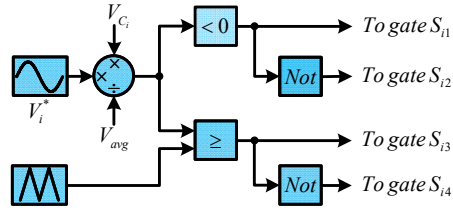

Figure 6: The proposed control method for the first abnormal situation.

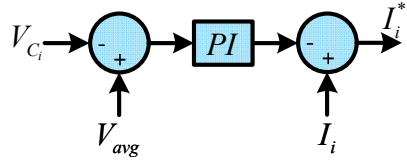

Figure 7: The proposed control method for limiting current reference.

There are no harmonics in the voltage that are multiples of three. Eq. (12) delivers for the maximum fundamental component of each phase voltage the proposed MMI can apply to the load

$$\begin{aligned} V_{i_{1rms}} &= \frac{2\sqrt{3}V_{DC}}{n\pi}\frac{1}{\sqrt{2}} \\ &= \frac{\sqrt{6}}{\pi}\frac{V_{DC}}{n} \cong \frac{0.78}{n}V_{DC} \cong 0.78 V_{C_i} \,. \end{aligned} \tag{13}$$

## 3.3 Additional Control Algorithms

Although the proposed MMI for multiphase machines can self-balance its capacitors, there are two situations in which an additional control algorithm is required. These situations are described, and appropriate control methods are proposed for the sake of the proposed MMI's generality.

### 3.3.1 If the load is not entirely in balance

For example, the load consists of different single-phase loads. Self-balancing will not work because the discussed electromagnetic coupling does not exist in this situation and the stable operation point of the circuit is not proper. The solution is straightforward in this case. As shown in Fig. 6, the modulation index (MI) should be controlled where $V_{C_i}$ is the $i$th capacitor's voltage and $V_{avg} = V_{DC}/n$. If the system is properly designed, occasional MI limitation will keep the capacitors balanced.

### 3.3.2 If it is necessary that the voltages are perfectly balanced

If a voltage imbalance occurs, the proposed topology will correct it, as previously stated in Section 3. However, due to the nature of the circuit, a small amount of voltage fluctuation (less than 2.5 percent) is natural. Additional control methods should be implemented if required to achieve lower voltage fluctuations. In addition to the previous method for manipulating the MI, two more techniques for controlling the fluctuation are proposed here, altering the machine control method. Different machine control methods are presented in the literature, such as vector control and direct torque control. The controlling parameter in each method should be limited. Fig. 7 shows the method to limit the current reference of $C_L$, Fig. 8 the procedure to determine the torque reference of the $C_L$ phase. $I_i$ and $T_i$ are the references produced by the machine control methods, and $I_i^*$ as well as $T_i^*$ are the limited references that should be considered in the rest of the machine control algorithm.

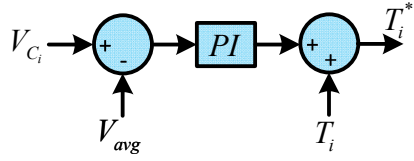

Figure 8: The proposed control method for limiting torque reference.

Table 2: System Parameters

| Parameters | Simulations | |
|---|---|---|
| | 5-level | 7-level |
| DC source voltage | 600 V | 400 V |
| SMs capacitance | 220 $\mu$F | |
| Capacitor resistance | 10 m$\Omega$ | |
| Switch model | MOSFET | |
| $R_DS$ | 10 m$\Omega$ | |
| $V_DS$ | - | |
| Electrical machine type | Induction | |
| Number of phases of machine | 3-phase | |
| Machine nominal voltage | 400 V | |
| Machine nominal power | 1 kW | |
| Rated speed | 1500 rpm | |

## 4 Simulation Verification

This section provides various simulation results in order to test the proposed MML's proper performance. The MAT-LAB/Simulink program served for the computer simulation with the parameters in Table 2. A three-phase induction machine is used to perform simulations for five-level and seven-level output voltages.

Fig. 9 and Fig. 10 present the steady-state simulation results of five- and seven-level inverters, respectively. To achieve the nominal voltage of the utilized machine (400 V), a 600 V DC voltage source for a five-level inverter or a 400 V DC voltage source for a seven-level inverter is required (according to Eq. (8)). The voltages of the capacitors fluctuate over time in a reasonable band and do not leave it, demonstrating their intrinsic balancing capability. The peak to peak voltage ripple of five-level and seven-level capacitors are approximately 4.4 V and 2.4 V, respectively, which are 2.2 % and 1.8 % fluctuation according to mean voltage. The currents of the capacitors are shown separately in the figures to demonstrate the system's proper performance with respect to the highest tolerable switch current.

Because the capacitors in the proposed MMI would be connected in parallel, smaller capacitors can be used in SMs, resulting in a significant reduction in total volume and cost.

Although steady-state measurements illustrate the stability of capacitor voltages in the $(N_L - 1)V_{DC}/2n$, the system has to be able to balance if a capacitor is severely unbalanced for any cause. Assume that in a system with five-level SMs, a capacitor significantly deviates from the mean voltage. Fig. 11 shows the system performance if the capacitor voltage is $300\,V$, which should be $200\,V$ in a balanced scenario. The system needed roughly 40 ms to balance the 100 V deviation, following which it resumed regular operation.

## 5 Conclusion

In this article, a unique MMI structure is presented, which combines multilevel and switched-capacitor features to drive a multiphase machine as an application example. With four switches and one capacitor, the core concept of the suggested topology may provide a three-level output voltage waveform. Seven switches and two self-balancing capacitors can be used to generate a five-level output. The SM architecture and the system design enable the output voltage to be increased or decreased compared to the input DC voltage. In addition, we presented closed-loop controller for particular loads, as well as the related modulation and control method. Finally, several simulation results verified the accurate performance of the suggested MMI.

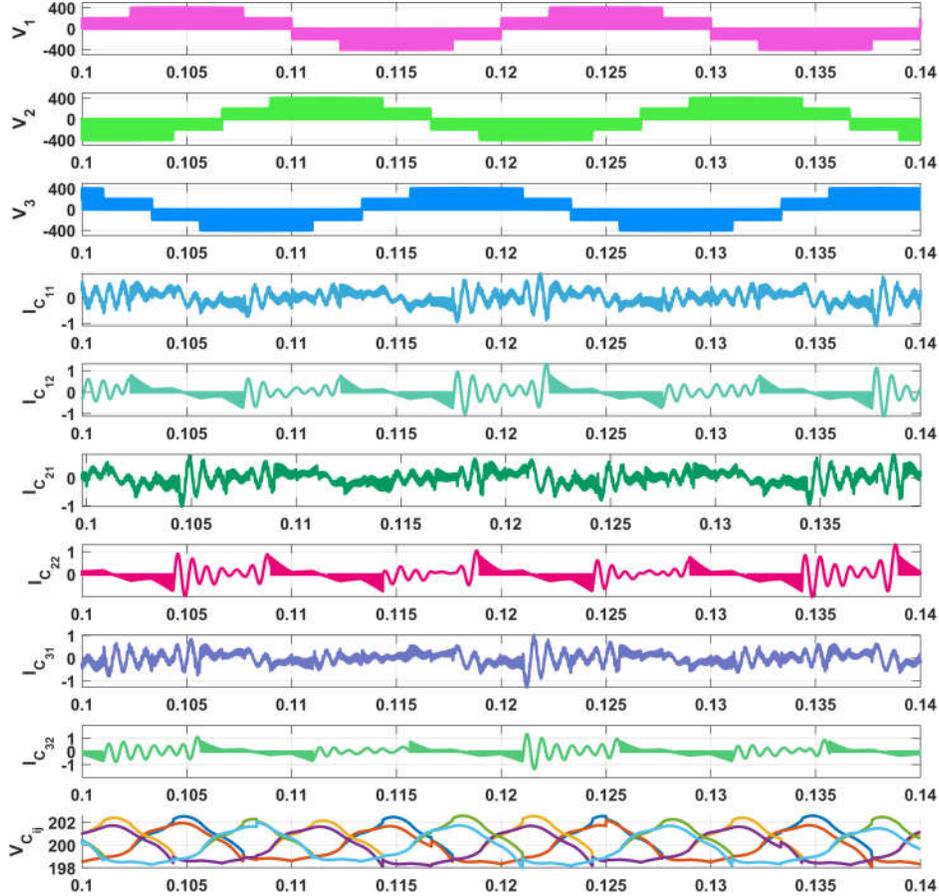

Figure 9: Steady-state simulation results for a five-level inverter; $V_1$ to $V_3$ are unfiltered five-level output voltage; $I_{C_{11}}$ to $I_{C_{32}}$ are currents of SMs capacitors; $V_{C_{ij}}$ are voltages of SMs capacitors.

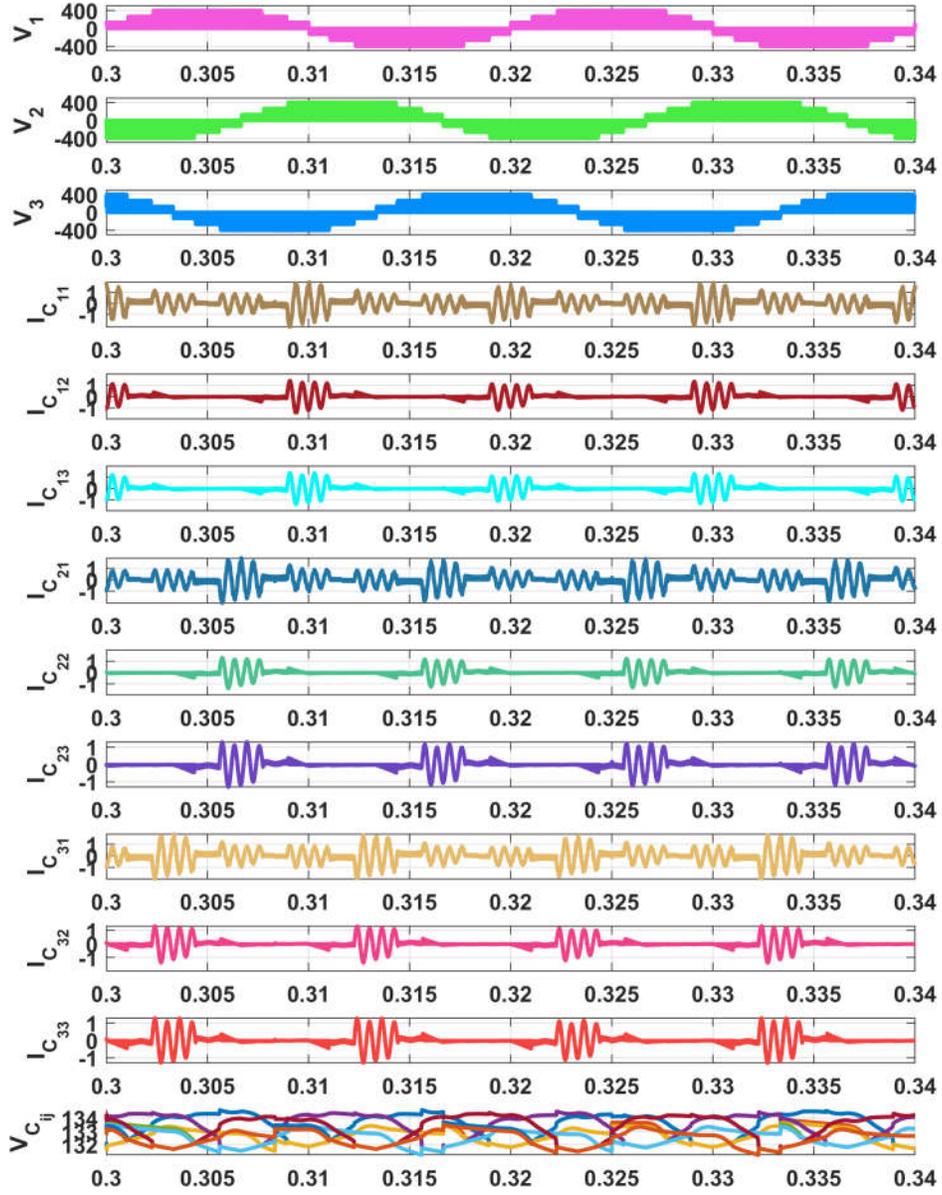

Figure 10: Steady state simulation results for seven-level inverter; $V_1$ to $V_3$ are unfiltered seven-level output voltage; $I_{C_{11}}$ to $I_{C_{33}}$ are currents of SMs capacitors; $V_{C_{ij}}$ are voltages of SMs capacitors.

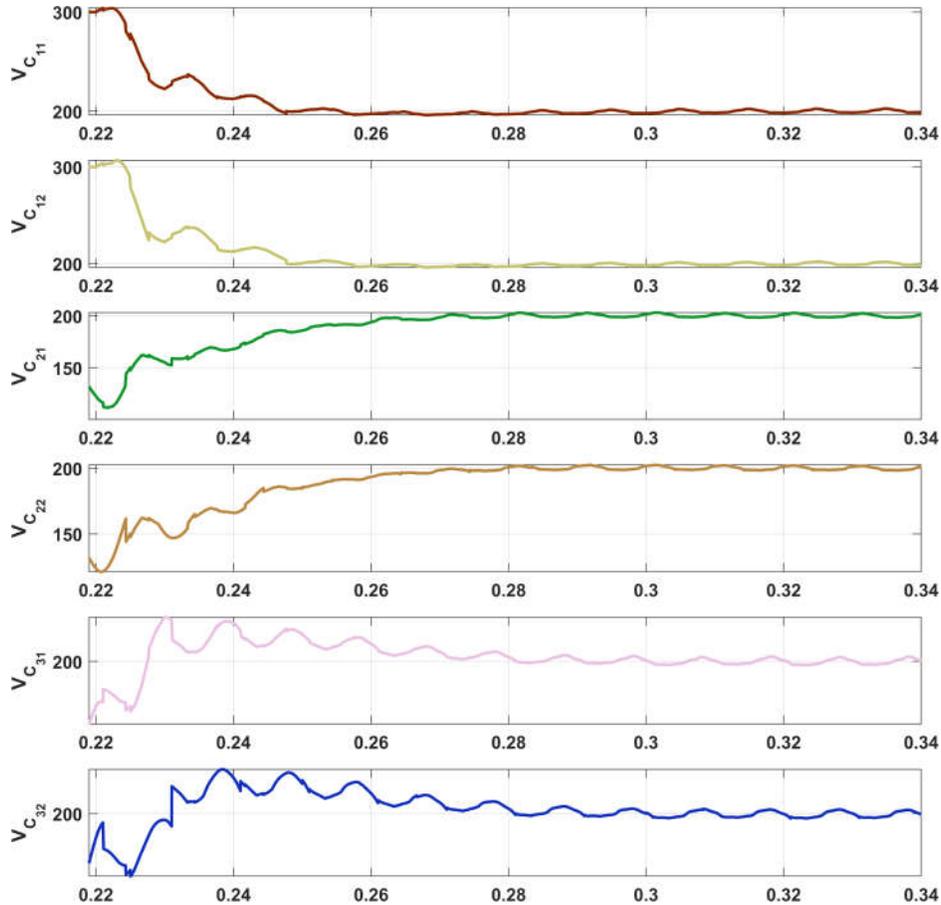

Figure 11: System performance in case of intentionally unbalancing capacitors, the system works effectively even if the capacitors are imbalanced.